\documentclass[preprint,superscriptaddress]{revtex4-1}

\usepackage{amssymb}
\usepackage{amsmath}
\usepackage{graphicx}
\usepackage{dcolumn}
\usepackage{bm}
\usepackage{verbatim}
\usepackage{setspace}
\usepackage{fancyhdr,lastpage}

\begin{document}

\newcommand{\fourqquads}{{\qquad \qquad \qquad \qquad}}
\newcommand{\eightqquads}{{\qquad \qquad \qquad \qquad \qquad \qquad \qquad \qquad}}
\newcommand{\fournegspc}{{\! \! \! \!}}
\newcommand{\eightnegspc}{{\! \! \! \! \! \! \! \!}}
\newcommand{\AGAt}{${\rm Al_{0.10}Ga_{0.90}As}$}
\newcommand{\AGAff}{${\rm Al_{0.55}Ga_{0.45}As}$}
\newcommand{\AGAy}{${\rm Al_{\it y}Ga_{1-{\it y}}As}$}
\newcommand{\AGAx}{${\rm Al_{\it x}Ga_{1-{\it x}}As}$}
\newcommand{\AGAf}{${\rm Al_{0.15}Ga_{0.85}As}$}
\newcommand{\IGAx}{${\rm In_{0.53}Ga_{0.47}As/In_{0.52}Al_{0.48}As}$}
\newcommand{\IGx}{${\rm In_{0.53}Ga_{0.47}As}$}
\newcommand{\AGA}{GaAs/AlGaAs}
\newcommand{\rmIR}{{\mathrm{IR}}}
\newcommand{\IGA}{InGaAs/InAlAs}
\newcommand{\WcmK}{W/cm$\cdot$K}
\newcommand{\E}[1]{$E_{#1}$}
\newcommand{\N}[1]{$N_{#1}$}
\newcommand{\DN}{$\Delta N_{32}$}
\newcommand{\taut}[2]{$\tau_{#1}^{#2}$}
\newcommand{\Schrodinger}{Schr\"{o}dinger}
\newcommand{\epmiwt}{{e^{\pm i \omega t}}}
\newcommand{\empiwt}{{e^{\mp i \omega t}}}
\newcommand{\eiwt}{{e^{i \omega t}}}
\newcommand{\epiwt}{{e^{+i \omega t}}}
\newcommand{\emiwt}{{e^{-i \omega t}}}
\newcommand{\tauul}{{\tau_{u\rightarrow l}}}
\newcommand{\hw}{{\hbar \omega}}
\newcommand{\hwz}{{\hbar \omega_0}}
\newcommand{\cc}{{\mathrm{c.c.}}}
\newcommand{\Ef}{{\mathcal{E}}}
\newcommand{\Pol}{{\mathcal{P}}}
\newcommand{\Rdiff}{{\mathcal{R}}}
\newcommand{\Rdiffm}{{\mathcal{R}_\mathrm{th}^-}}
\newcommand{\Rdiffp}{{\mathcal{R}_\mathrm{th}^+}}
\newcommand{\dw}{{d \omega}}
\newcommand{\hwab}{{\hbar \omega_{a b}}}
\newcommand{\Np}{{N_\mathrm{p}}}
\newcommand{\Pout}{{P_\mathrm{out}}}
\newcommand{\GammabyNp}{{\frac{\Gamma}{\Np}}}
\newcommand{\etaIth}{{\eta_{I\mathrm{th}}}}
\newcommand{\etaI}{{\eta_I}}
\newcommand{\nst}{{n^\mathrm{st}_\mathrm{ph}}}
\newcommand{\nsp}{{n^\mathrm{sp}_\mathrm{ph}}}
\newcommand{\nph}{{n_\mathrm{ph}}}
\newcommand{\tpara}{{\tau_{3,\mathrm{para}}}}
\newcommand{\tsp}{{\tau_\mathrm{sp}}}
\newcommand{\tph}{{\tau_\mathrm{ph}}}
\newcommand{\tpsp}{{\tau'_\mathrm{sp}}}
\newcommand{\tst}{{\tau_\mathrm{st}}}
\newcommand{\gmat}{{g_{\mathrm{mat}}}}
\newcommand{\Ith}{{I_{\mathrm{th}}}}
\newcommand{\Ithp}{{I_{\mathrm{th}}^+}}
\newcommand{\Ithm}{{I_{\mathrm{th}}^-}}
\newcommand{\Jth}{{J_{\mathrm{th}}}}
\newcommand{\Jthf}{{J_{\mathrm{th},5 \mathrm{K}}}}
\newcommand{\Jmaxf}{{J_{\mathrm{max},5 \mathrm{K}}}}
\newcommand{\Jtht}{{J_{\mathrm{th},10 \mathrm{K}}}}
\newcommand{\Jmaxt}{{J_{\mathrm{max},10 \mathrm{K}}}}
\newcommand{\Tactive}{{T_{\mathrm{active}}}}
\newcommand{\Tcw}{{T_{\mathrm{cw}}}}
\newcommand{\fosc}{{f_{\mathrm{osc}}}}
\newcommand{\Tmax}{{T_{\mathrm{max}}}}
\newcommand{\Tmaxp}{{T_{\mathrm{max,pul}}}}
\newcommand{\Tmaxc}{{T_{\mathrm{max,cw}}}}
\newcommand{\Jmax}{{J_{\mathrm{max}}}}
\newcommand{\Jpeak}{{J_{\mathrm{peak}}}}
\newcommand{\Imax}{{I_{\mathrm{max}}}}
\newcommand{\Vac}{{V_{\mathrm{ac}}}}
\newcommand{\Vmod}{{V_{\mathrm{mod}}}}
\newcommand{\Aac}{{A_{\mathrm{ac}}}}
\newcommand{\Vcav}{{V_{\mathrm{cav}}}}
\newcommand{\Acav}{{A_{\mathrm{cav}}}}
\newcommand{\SEQUAL}{\mbox{SEQUAL}}
\newcommand{\mez}{{m^{\ast}(z)}}
\newcommand{\me}{m^\ast}
\newcommand{\e}{m^{\ast}(z,E)}
\newcommand{\mezei}{m^{\ast}(z,E_i)}
\newcommand{\mezef}{m^{\ast}(z,E_f)}
\newcommand{\ddz}{\frac{\partial}{\partial z}}
\newcommand{\pddt}{\frac{\partial}{\partial t}}
\newcommand{\ddt}{\frac{d}{d t}}
\newcommand{\etal}{{\it {et~al.\ }}}
\newcommand{\ie}{\mbox{i.\ e.}}
\newcommand{\psicz}{\psi_c(z)}
\newcommand{\psiciz}{\psi_c^{(i)}(z)}
\newcommand{\psicfz}{\psi_c^{(f)}(z)}
\newcommand{\epsc}{\epsilon_{\mathrm{core}}}
\newcommand{\epsmc}{\epsilon_{\mathrm{m,core}}}
\newcommand{\epsr}{\epsilon_{\mathrm{r}}}
\newcommand{\epsm}{\epsilon_{\mathrm{m}}}
\newcommand{\epsd}{\epsilon_{\mathrm{d}}}
\newcommand{\epsdt}{\epsilon^2_{\mathrm{d}}}
\newcommand{\epsz}{\epsilon_{0}}
\newcommand{\km}{k_\mathrm{m}}
\newcommand{\kd}{k_\mathrm{d}}
\newcommand{\kdt}{k^2_\mathrm{d}}
\newcommand{\degC}[1]{{#1\,^\circ\mathrm{C}}}
\newcommand{\betaz}{{\beta_z}}
\newcommand{\neff}{{n_\mathrm{eff}}}
\newcommand{\wpl}{{\omega_{\mathrm{p}}}}
\newcommand{\wplt}{{\omega^2_{\mathrm{p}}}}
\newcommand{\wspl}{{\omega_{\mathrm{sp}}}}
\newcommand{\kpari}{{\mathbf{k}_{\|,i}}}
\newcommand{\kparf}{{\mathbf{k}_{\|,f}}}
\newcommand{\kparm}{{\mathbf{k}_{\|,m}}}
\newcommand{\kparn}{{\mathbf{k}_{\|,n}}}
\newcommand{\kpara}{{\mathbf{k}_{\|,a}}}
\newcommand{\kparb}{{\mathbf{k}_{\|,b}}}
\newcommand{\ikpari}{{i,{\mathbf k}_{\|,i}}}
\newcommand{\fkparf}{{f,{\mathbf k}_{\|,f}}}
\newcommand{\mkparm}{{m,{\mathbf k}_{\|,m}}}
\newcommand{\akpara}{{a,{\mathbf k}_{\|,a}}}
\newcommand{\bkparb}{{b,{\mathbf k}_{\|,b}}}
\newcommand{\phinr}{{\phi_n \rangle}}
\newcommand{\phiml}{{\langle \phi_m}}
\newcommand{\psitket}{{|\psi(t)\rangle}}
\newcommand{\iket}{{| i \rangle}}
\newcommand{\jket}{{| j \rangle}}
\newcommand{\aket}{{| a \rangle}}
\newcommand{\bket}{{| b \rangle}}
\newcommand{\ket}[1]{{| #1 \rangle}}
\newcommand{\bra}[1]{{\langle #1 |}}
\newcommand{\ibra}{{\langle i |}}
\newcommand{\jbra}{{\langle j |}}
\newcommand{\abra}{{\langle a |}}
\newcommand{\bbra}{{\langle b |}}
\newcommand{\kpar}{{\mathbf{k}_{\|}}}
\newcommand{\ahat}{{\hat{a}}}
\newcommand{\Vbias}{{V_\mathrm{bias}}}
\newcommand{\ahatc}{{\hat{a}^\dagger}}
\newcommand{\eikr}{{e^{i \mathbf{k}\cdot\mathbf{r}}}}
\newcommand{\emikr}{{e^{-i \mathbf{k}\cdot\mathbf{r}}}}
\newcommand{\br}{{\mathbf{r}}}
\newcommand{\bz}{{\mathbf{z}}}
\newcommand{\bk}{{\mathbf{k}}}
\newcommand{\bu}{{\mathbf{u}}}
\newcommand{\usigma}{{\mathbf{u}_\sigma}}
\newcommand{\ksigma}{{\mathbf{k},\sigma}}
\newcommand{\rpar}{{\mathbf{r}_\|}}
\newcommand{\bA}{{\mathbf A}}
\newcommand{\bE}{{\mathbf E}}
\newcommand{\ki}{{\mathbf{k}_i}}
\newcommand{\kf}{{\mathbf{k}_f}}
\newcommand{\kj}{{\mathbf{k}_j}}
\newcommand{\kg}{{\mathbf{k}_g}}
\newcommand{\qpara}{{\bf q}_{\parallel}}
\newcommand{\qz}{{\bf q}_z}
\newcommand{\LO}{{\mathrm{LO}}}
\newcommand{\hwLO}{{\hbar\omega_{\mathrm{LO}}}}
\newcommand{\wLO}{{\omega_{\mathrm{LO}}}}
\newcommand{\ELO}{{E_{\mathrm{LO}}}}
\newcommand{\bracomboa}{{| n_a^\ksigma : \akpara \rangle}}
\newcommand{\bracombob}{{| n_b^\ksigma : \bkparb \rangle}}
\newcommand{\comboa}{{n_a^\ksigma : \akpara}}
\newcommand{\combob}{{n_b^\ksigma : \bkparb}}
\newcommand{\mat}{{\mathrm{mat}}}
\newcommand{\rmc}{{\mathrm{c}}}
\newcommand{\rms}{{\mathrm{s}}}
\newcommand{\rmmin}{{\mathrm{min}}}
\newcommand{\rmmax}{{\mathrm{max}}}
\newcommand{\rmph}{{\mathrm{ph}}}
\newcommand{\rmps}{{\mathrm{ps}}}
\newcommand{\rmst}{{\mathrm{st}}}
\newcommand{\rmsp}{{\mathrm{sp}}}
\newcommand{\ghz}{{\mathrm{GHz}}}
\newcommand{\thz}{{\mathrm{THz}}}
\newcommand{\meV}{{\mathrm{meV}}}
\newcommand{\eV}{{\mathrm{eV}}}
\newcommand{\ps}{{\mathrm{ps}}}
\newcommand{\rmand}{{\mathrm{and}}}
\newcommand{\rmwhere}{{\mathrm{where}}}
\newcommand{\rmpara}{{\mathrm{para}}}
\newcommand{\cav}{{\mathrm{cav}}}
\newcommand{\thD}{{\mathrm{3D}}}
\newcommand{\twD}{{\mathrm{2D}}}
\newcommand{\rmeff}{{\mathrm{eff}}}
\newcommand{\rmhot}{{\mathrm{hot}}}
\newcommand{\rmLO}{{\mathrm{LO}}}
\newcommand{\rmmW}{{\mathrm{mW}}}
\newcommand{\rmuW}{{\mu\mathrm{W}}}
\newcommand{\rmK}{{\mathrm{K}}}
\newcommand{\nGaAs}{{n_\mathrm{GaAs}}}
\newcommand{\nSi}{{n_\mathrm{Si}}}
\newcommand{\rmGaAs}{{\mathrm{GaAs}}}
\newcommand{\slopeeff}{{\frac{\alpha_{\rmm,\rmf}}{\alpha_\rmw+\alpha_{\rmm,\rmf}+\alpha_{\rmm,\rmr}}}}
\newcommand{\rmfor}{{\mathrm{for}}}
\newcommand{\inlimit}{{\mathrm{in \ the \ limit}}}
\newcommand{\Dinj}{{\Delta_\mathrm{inj}}}
\newcommand{\Dcol}{{\Delta_\mathrm{col}}}
\newcommand{\rmtb}{{\mathrm{tb}}}
\newcommand{\rmr}{{\mathrm{r}}}
\newcommand{\rmf}{{\mathrm{f}}}
\newcommand{\rme}{{\mathrm{e}}}
\newcommand{\rmR}{{\mathrm{R}}}
\newcommand{\rmi}{{\mathrm{i}}}
\newcommand{\rmnm}{{\mathrm{nm}}}
\newcommand{\rmu}{{\mathrm{u}}}
\newcommand{\rml}{{\mathrm{l}}}
\newcommand{\rmj}{{\mathrm{j}}}
\newcommand{\rmm}{{\mathrm{m}}}
\newcommand{\rmw}{{\mathrm{w}}}
\newcommand{\rmmm}{{\mathrm{mm}}}
\newcommand{\rmmat}{{\mathrm{mat}}}
\newcommand{\rmth}{{\mathrm{th}}}
\newcommand{\rmV}{{\mathrm{V}}}
\newcommand{\defas}{{\stackrel{\triangle}{=}}}
\newcommand{\itof}{{i \rightarrow f}}
\newcommand{\ftoi}{{f \rightarrow i}}
\newcommand{\atob}{{a \rightarrow b}}
\newcommand{\btoa}{{b \rightarrow a}}
\newcommand{\qsig}{{{\bf q},\sigma}}
\newcommand{\epol}{{\bf \hat{e}}_{{\bf q},\sigma}}
\newcommand{\zhat}{{\bf \hat{z}}}
\newcommand{\xhat}{{\bf \hat{x}}}
\newcommand{\pop}{{\hat{\mathbf{p}}}}
\newcommand{\rop}{{\hat{\mathbf{r}}}}
\newcommand{\rzop}{{\hat{\mathbf{z}}}}
\newcommand{\rpop}{{\hat{\mathbf{r}}_{\|}}}
\newcommand{\Aop}{{\hat{\mathbf{A}}}}
\newcommand{\Eop}{{\hat{\mathbf{E}}}}
\newcommand{\Hop}{{\hat{H}}}
\newcommand{\Oop}{{\hat{O}}}
\newcommand{\Omat}{{\bar{\bar{O}}}}
\newcommand{\Htbop}{{\hat{H}_\mathrm{tb}}}
\newcommand{\Htbmat}{{\bar{\bar{H}}_\mathrm{tb}}}
\newcommand{\Hext}{{\hat{H}}}
\newcommand{\Hextop}{{\hat{H}_\mathrm{ext}}}
\newcommand{\Hextmat}{{\bar{\bar{H}}_\mathrm{ext}}}
\newcommand{\Hmat}{{\bar{\bar{H}}}}
\newcommand{\rhomat}{{\bar{\bar{\rho}}}}
\newcommand{\rhoop}{{\hat{\rho}}}
\newcommand{\ntot}{{n_\mathrm{tot}}}
\newcommand{\Lmod}{{L_\mathrm{mod}}}
\newcommand{\Lcav}{{L_\mathrm{cav}}}
\newcommand{\tulhotLO}{{\tau^\rmhot_{ul,\rmLO}}}
\newcommand{\tulLO}{{\tau_{ul,\rmLO}}}
\newcommand{\tFLparb}{{\tau_{5\rightarrow(6,3,2,1)}}}
\newcommand{\tFLpar}{{\tau_{5,\mathrm{par}}}}
\newcommand{\Tpure}{{T^*_2}}
\newcommand{\td}{{\tau_{\|}}}
\newcommand{\tdgen}[1]{{\tau_{\|#1}}}
\newcommand{\tdtw}{{\tau_{\|,2}}}
\newcommand{\tdtwp}{{\tau_{\|,2'}}}
\newcommand{\tdtwpsq}{{\tau^2_{\|,2'}}}
\newcommand{\tdtwth}{{\tau_{\|,23}}}
\newcommand{\tdtwpth}{{\tau_{\|,2'3}}}
\newcommand{\tdth}{{\tau_{\|,3}}}
\newcommand{\tdsq}{{\tau^2_{\|}}}
\newcommand{\tdsqgen}[1]{{\tau^2_{\|,#1}}}
\newcommand{\Ahsqgen}[1]{{\left(\frac{\Delta_{#1}}{\hbar}\right)^2}}
\newcommand{\Ahsqgenb}[1]{{\Omega_{#1}^2}}
\newcommand{\Ahsq}{{\left(\frac{\Delta_0}{\hbar}\right)^2}}
\newcommand{\Ahcsq}{{\left(\frac{\delzc}{\hbar}\right)^2}}
\newcommand{\Ehsqgen}[1]{{\left(\frac{E_{#1}}{\hbar}\right)^2}}
\newcommand{\Ehsq}{{\left(\frac{E_{1'3}}{\hbar}\right)^2}}
\newcommand{\Ehcsq}{{\left(\frac{E_{22'}}{\hbar}\right)^2}}
\newcommand{\delop}{{\hat{\nabla}}}
\newcommand{\zzif}{|z_{i \rightarrow f}|^2}
\newcommand{\deltakr}{{\delta^{\mathrm{kr}}}}
\newcommand{\delnu}{{\Delta \nu}}
\newcommand{\delfwhm}{{\Delta \nu_\mathrm{FWHM}}}
\newcommand{\Erad}{{E_\mathrm{rad}}}
\newcommand{\frad}{{f_\mathrm{rad}}}
\newcommand{\zrad}{{z_\mathrm{rad}}}
\newcommand{\nurad}{{\nu_\mathrm{rad}}}
\newcommand{\delec}{{\Delta E_\mathrm{c}}}
\newcommand{\delnth}{{\Delta n_\mathrm{th}}}
\newcommand{\drr}{{\Delta \mathcal{R}_\mathrm{th}/\mathcal{R}_\mathrm{th}}}
\newcommand{\drrfrac}{{\frac{\Delta \mathcal{R}_\mathrm{th}}{\mathcal{R}_\mathrm{th}}}}
\newcommand{\Rth}{{\mathcal{R}_\mathrm{th}}}
\newcommand{\delzc}{{\Delta^{\mathrm{c}}_0}}
\newcommand{\gth}{{g_\mathrm{th}}}
\newcommand{\delN}{{\Delta N}}
\newcommand{\icm}{{{\mathrm{cm}}^{-1}}}
\newcommand{\Appcm}{{{\mathrm{A}/\mathrm{cm}}^{2}}}
\newcommand{\iicm}{{{\mathrm{cm}}^{-2}}}
\newcommand{\iiicm}{{{\mathrm{cm}}^{-3}}}
\newcommand{\iiium}{{{\mu\mathrm{m}}^{-3}}}
\newcommand{\iiim}{{{\mathrm{m}}^{-3}}}
\newcommand{\kacmm}{{\rm kA/cm}^2}
\newcommand{\real}[1]{{\mathcal{R}\mathrm{e}\{#1\}}}
\newcommand{\imag}[1]{{\mathcal{I}\mathrm{m}\{#1\}}}
\newcommand{\um}{{\mu\mathrm{m}}}
\newcommand{\IV}{{$I$-$V$}}
\newcommand{\VIs}{{$V$-$I$s}}
\newcommand{\IVs}{{$I$-$V$s}}
\newcommand{\VI}{{$V$-$I$}}
\newcommand{\LI}{{$L$-$I$}}
\newcommand{\LIs}{{$L$-$I$s}}
\newcommand{\LVs}{{$L$-$V$s}}
\newcommand{\PI}{{$P$-$I$}}
\newcommand{\LV}{{$L$-$V$}}
\newcommand{\RV}{{$\mathcal{R}$-$V$}}
\newcommand{\RVs}{{$\mathcal{R}$-$V$s}}
\newcommand{\RI}{{$\mathcal{R}$-$I$}}
\newcommand{\RIs}{{$\mathcal{R}$-$I$s}}
\newcommand{\JV}{{$J$-$V$}}
\newcommand{\GV}{{$G$-$V$}}
\newcommand{\CV}{{$C$-$V$}}
\newcommand{\IB}{{$I$-$B$}}
\newcommand{\GB}{{$G$-$B$}}
\newcommand{\dVdI}{{$dV/dI$}}
\newcommand{\dVdIV}{{$dV/dI$-$V$}}
\newcommand{\dVdII}{{$dV/dI$-$I$}}
\newcommand{\ebar}{\overline{\eta}}
\newcommand{\Perot}{{P\'{e}rot}}
\newcommand{\phos}{{$\mathrm{H_3PO_4:H_2O_2:H_2O}$}}
\newcommand{\sulf}{{$\mathrm{H_2SO_4:H_2O_2:H_2O}$}}
\newcommand{\amm}{{$\mathrm{NH_4OH:H_2O_2:H_2O}$}}
\newcommand{\ammlap}{{$\mathrm{NH_4OH:H_2O_2}$}}
\newcommand{\alumina}{{$\mathrm{Al_2O_3}$}}
\newcommand{\alcoat}{{$\mathrm{Al_2O_3/Ti/Au/Al_2O_3}$}}
\newcommand{\ang}{{\mathrm{\AA}}}

\title{Large static tuning of narrow-beam terahertz plasmonic lasers operating at $78~\rmK$}



\author{Chongzhao Wu}
\affiliation{Department of Electrical and Computer Engineering, Lehigh University, Bethlehem, PA 18015, USA}

\author{Yuan Jin}
\affiliation{Department of Electrical and Computer Engineering, Lehigh University, Bethlehem, PA 18015, USA}

\author{John L. Reno}
\affiliation{Sandia National Laboratories, Center of Integrated Nanotechnologies, MS 1303, Albuquerque, NM 87185, USA}

\author{Sushil Kumar}
\email[Email: ]{sushil@lehigh.edu}
\affiliation{Department of Electrical and Computer Engineering, Lehigh University, Bethlehem, PA 18015, USA}


\begin{abstract}
A new tuning mechanism is demonstrated for single-mode metal-clad plasmonic lasers, in which refractive-index of the laser's
surrounding medium affects the resonant-cavity mode in the same vein as refractive-index of gain medium inside the
cavity. Reversible, continuous, and mode-hop-free tuning of $\sim 57~\ghz$ is realized for single-mode narrow-beam
terahertz plasmonic quantum-cascade lasers (QCLs), 
which is demonstrated at much more practical temperature of $78~\rmK$.
The tuning is based on post-process deposition/etching of a dielectric (Silicon-dioxide) on a QCL chip that has
already been soldered and wire-bonded onto a copper mount. This is a considerably larger tuning range compared to
previously reported results for terahertz QCLs with directional far-field radiation patterns. 
The key enabling mechanism for tuning
is a recently developed antenna-feedback scheme for plasmonic lasers, which leads to generation of hybrid
surface-plasmon-polaritons propagating outside the cavity of the laser with a large spatial extent.  The effect
of dielectric deposition on QCL's characteristics is investigated in detail including that on maximum operating
temperature, peak output power and far-field radiation patterns.  
Single-lobed beam with low divergence ($<7^\circ$) is maintained through the tuning range.
The antenna-feedback scheme is ideally suited for modulation of plasmonic lasers and their sensing applications due to the
sensitive dependence of spectral and radiative properties of the laser on its surrounding medium.

\end{abstract}


\flushbottom
\maketitle
\thispagestyle{empty}

\section*{Introduction}

Plasmonic lasers that utilize metal-clad cavities confine electromagnetic energy in the form of
surface-plasmon-polaritons (SPPs) at subwavelength dimensions. They have seen rapid development for targeted
applications in nanoscale optics and integrated optical sensing. The most common types of plasmonic lasers utilize
Fabry-\Perot\ type cavities in which one dimension is not subwavelength and is typically much longer than the other two
dimensions~\cite{hill:mimlaser,oulton:plasmonlaser,lu:nanolaser}. For applications requiring spectral purity
(single-mode operation) at the desired wavelength, tunability is a desired characteristic that is challenging to
implement for such lasers. This stems from the difficulty in external control of the resonant-cavity mode such as that
in conventional external-cavity tuning techniques for semiconductor lasers, since the subwavelength characteristics of
such lasers lead to poor coupling between the cavity-modes with free-space propagating modes. Prior work on development
of tunable plasmonic lasers has primarily focused on changing the gain medium of the cavity
itself~\cite{lu:tunable,yang:tunable}. Terahertz quantum-cascade lasers
(QCLs)~\cite{kohler:laser,williams:review,vitiello:reviewQCLs} with metallic cavities~\cite{williams:metal} are one such
subset of plasmonic lasers, which also suffer for such challenges with respect to their tunability.

Terahertz QCLs are the brightest available solid-state sources~\cite{chattopadhyay:thzsources} of coherent terahertz
radiation. They have witnessed significant development in the past decade in multiple areas such as wave engineering
with distributed-feedback (DFB) techniques~\cite{sirtori:review13}, output-power~\cite{li:1wpower}, temperature
performance, frequency-coverage~\cite{scalari:review,kumar:review10,fathololoumi:thzqcl200K}, and frequency
tunability~\cite{vitiello:tunable}. Single-mode terahertz QCLs are particularly required for a multitude of applications
in sensing and spectroscopy. However, due to limits on accuracy of lithography utilized to implement DFB, it is still
challenging to achieve single-mode emission from DFB QCLs at the exact desired frequency for applications such as
high-resolution heterodyne spectroscopy~\cite{ren:heterodyne,hubers:THzspectroscopy}. Frequency tuning post-fabrication
is desired to overcome this problem.  Up to now, a variety of tuning techniques for terahertz QCLs have been
demonstrated. Temperature tuning is simple but lacks large tunability~\cite{kumar:surfemit}. Techniques such as applying
external-cavities~\cite{xu:tunable,lee:tunable} or modal perturbation using electromechanical
methods~\cite{qin:tunableMEMS,han:broad,castellano:tuning} have the ability to achieve large tuning range but at the
cost of discontinuous tuning, or system complexity and poor radiation patterns and temperature performance for such QCLs
respectively.

A static-tuning technique based on post-process dielectric deposition was recently demonstrated for terahertz
QCLs with third-order DFB operating at $\sim 10~\rmK$ with a tuning of $\sim 5~\ghz$~\cite{turcinkova:tunableqcl}. 
(The tuning could be extended while the QCL is in operation, but only temporarily, by deposition of solid nitrogen
through multiple-cycle condensation to $\sim 25~\ghz$ in a liquid-Helium cryostat).
Along similar lines, we here demonstrate a greatly enhanced continuous tuning range of $\sim 57~\ghz$ for single-mode
QCLs emitting at $2.8~\thz$ and operating at $78~\rmK$, achieved by deposition of Silicon-dioxide post-fabrication, and
with significantly improved beam profiles. Prior work on large ($>20~\ghz$) tuning in
Refs.~[\onlinecite{qin:tunableMEMS,han:broad,turcinkova:tunableqcl}] is achieved by perturbing the lateral evanescent
mode of metallic cavities, which requires the cavities to be made very narrow (deep-subwavelength, $\lesssim 10~\um$
wide cavities with the standard $10~\um$ thickness), and thereby makes the fabrication very challenging (with the
requirement of sophisticated dry-etching techniques) as well as degrades the temperature and power performance of the
QCLs considerably. In contrast, the tuning in this report places no such restrictions on the dimensions of the QCL
cavity and is demonstrated for wide cavities processed with routine wet-etching methods. The large tuning is facilitated
by a recently developed ``antenna-feedback'' scheme for plasmonic lasers, which was experimentally implemented for
terahertz QCLs with metallic cavities~\cite{wu:feedback}. This mode-hop free tuning performance can be controlled by the
thickness of deposited Silicon-dioxide, is reversible, and can be implemented on already mounted and soldered QCL chips,
which makes it attractive for future commercialization of single-mode terahertz QCLs. Importantly, in contrast to all
previous tuning results for terahertz QCLs that have been demonstrated at temperatures close-to that of liquid-Helium,
this work achieves large tuning for QCLs operating in a liquid-Nitrogen cooled dewar since the tuning mechanism does
not impact the gain of the QCL sensitively. Neither does the tuning method mandate a large dynamic range in current for
the QCL in operation, which is in contrast with the methods relying on frequency-pulling due to Stark-shifted gain
spectrum with changing electrical bias of the QCL that requires operation at
low-temperatures~\cite{zhang:phocrystal,dunbar:microcavity,turv:elec}.

\begin{figure*}[htbp]
\centering
\includegraphics[width=4.0in]{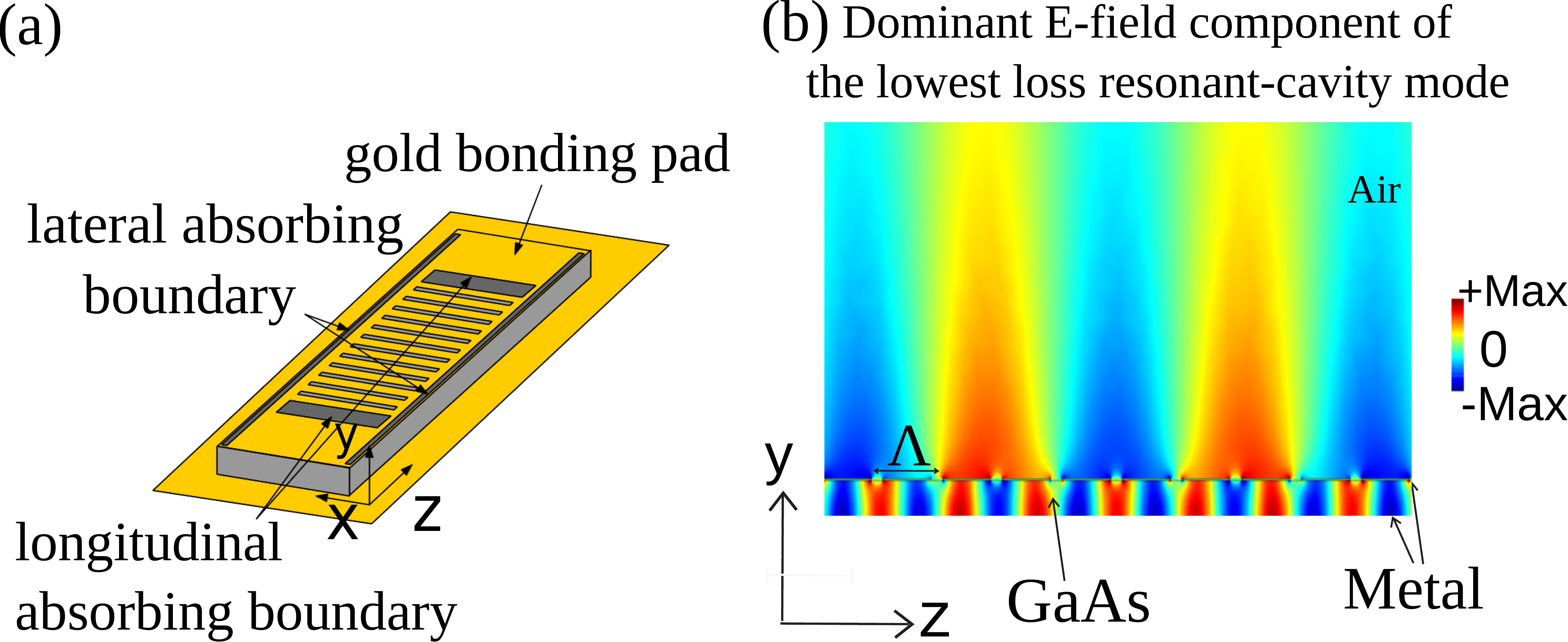}
\caption{
(a) 
Schematic of a terahertz plasmonic laser cavity with the antenna-feedback scheme. A parallel-plate
metallic cavity is shown with slit-like apertures in its top metal cladding with a specific
periodicity $\Lambda$ as determined from equation~\eqref{eq1}, which lead to coupling of a guided SPP wave inside the cavity
(interacting with the gain medium) with a single-sided SPP wave with a large spatial extent in the surrounding medium.
The lateral and longitudinal absorbing boundaries are implemented to selectively excite the lowest order
lateral and longitudinal resonant-cavity modes respectively.
(b)~The dominant electric-field component
($E_y$) of the lowest-loss resonant-cavity mode is plotted along $z$ axis, computed with finite-element (FEM)
simulations using a commercial software package (Comsol 4.3). The height of the cavity is $10~\um$. A hybrid SPP mode is excited
in the surrounding medium that is phase-locked to a guided SPP mode inside the cavity. 
}
\label{Fig1}
\end{figure*}

\section*{Description of the tuning technique}

Plasmonic lasers, of which the terahertz QCLs with metallic cavities are a specific example, lead to highly divergent
far-field radiation patterns owing to the subwavelength dimensions of their radiating apertures.  The antenna-feedback
scheme for plasmonic lasers~\cite{wu:feedback} couples the resonant surface-plasmon-polariton (SPP) mode in the cavity
to a highly directional far-field radiation pattern. This occurs by exciting hybrid SPPs with a large spatial extent in
the surrounding medium of the cavity by the mechanism of Bragg diffraction, which is affected by implementation of a
grating in the metal-cladding of the plasmonic laser as illustrated in Fig.~\ref{Fig1}(a). With the choice of a specific
periodicity in the grating, the resonant SPP wave inside the cavity is coupled via distributed-feedback (DFB) to a
hybrid SPP wave that can propagate in the surrounding medium on the opposite side of the metal-film being used in the
cavity of the plasmonic laser. 

For terahertz QCLs with metallic cavities the antenna-feedback scheme could be implemented by
introducing slit-like-apertures in its top metallic cladding as shown schematically in Fig.~\ref{Fig1}(b). 
For a chosen grating periodicity $\Lambda$, a DFB mode with free-space wavelength $\lambda$ (and frequency
$\nu=c/\lambda$) given by the equation
\begin{eqnarray}
\lambda & = & (n_\rmc + n_\rms) \; \Lambda
\label{eq1}
\end{eqnarray}
is resonantly excited for the case when the coupling between the SPP wave in the cavity and the SPP
wave in the surrounding medium is via first-order Bragg diffraction~\cite{wu:feedback}.
Here, $n_\rmc$ is the effective propagation index of the SPP wave inside the cavity (which is
approximately same as the refractive-index of the cavity's active medium) and $n_\rms$ is the effective propagation
index of the hybrid SPP wave in the surrounding medium (which is approximately same as the refractive-index of the
surrounding medium).

Equation~\eqref{eq1} is distinctly different from that of any other solid-state DFB lasers that has been reported in
literature, in that the refractive index of the surrounding medium is equally important in setting the
resonant-frequency of the DFB mode of the laser as is the refractive index of the active medium inside the cavity.  In
other conventional solid-state lasers, the effect of the surrounding medium on the resonant-frequency is primarily a
function of the fraction of evanescent mode of the cavity that propagates in its surroundings (typically represented by
a mode-confinement factor, $\Gamma$), which is also the case for the tunable terahertz QCL reported in
Ref.~\onlinecite{turcinkova:tunableqcl}.  However, in the antenna-feedback scheme, the contribution of the surrounding
medium is due to considerations of phase-matching~\cite{wu:feedback} rather than the fraction of the overall
electromagnetic energy propagating outside the cavity. Consequently, the resonant-frequency of the DFB mode depends
sensitively on the refractive-index of the surrounding medium that could be altered to tune the frequency of the
plasmonic laser, which is the technique employed for terahertz QCLs presented here. To the best of our knowledge, this
is a unique mechanism to tune the frequency of a solid-state laser for which there is no analogous precedent in
literature.

\begin{figure*}[htbp]
\centering
\includegraphics[width=5.5in]{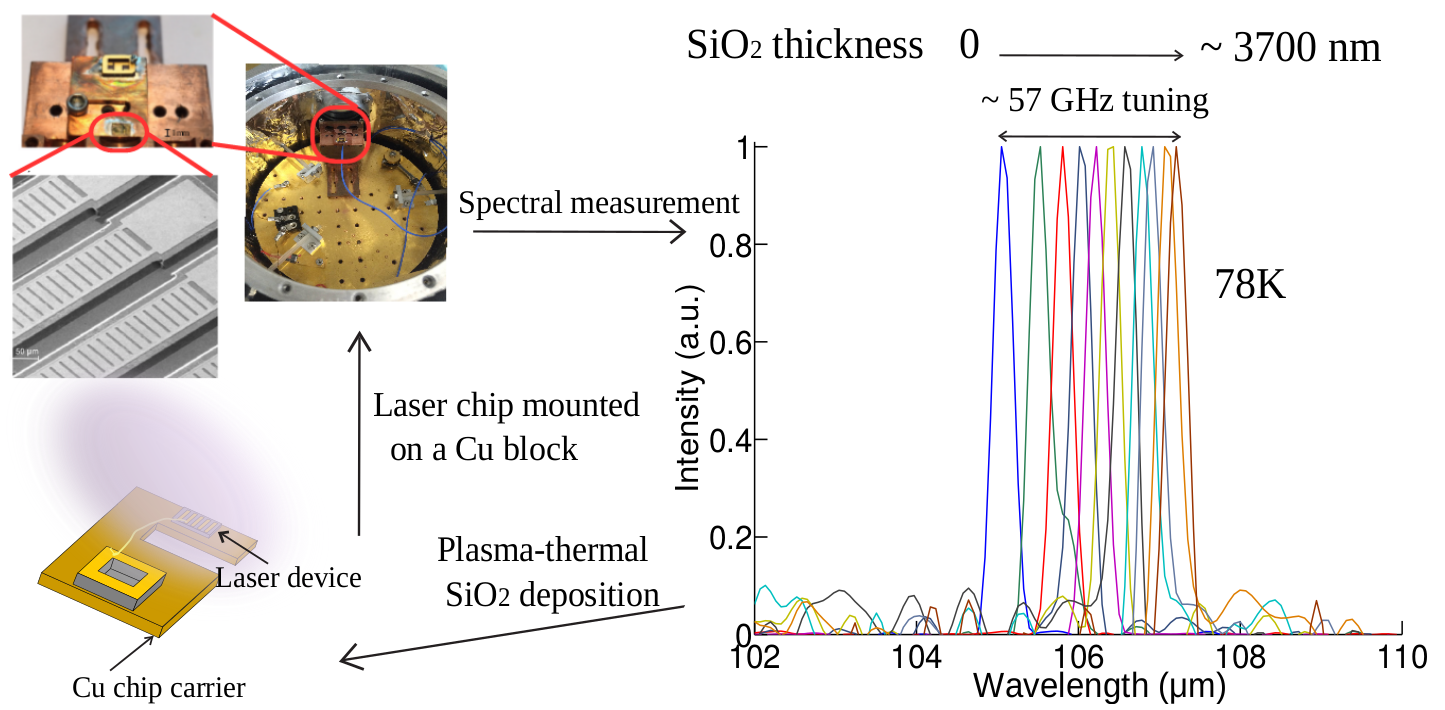}
\caption{
Optical image of a mounted QCL semiconductor chip out of which one of the QCLs with the antenna-feedback scheme is
wire-bonded for electrical characterization. The QCL chip is soldered on a small copper chip that itself is screwed onto
a bigger copper heat-sink to be mounted on the cold-plate of a cryogenic dewar. Multiple rounds of Silicon-dioxide
deposition and cryogenic measurements were performed on the soldered and mounted QCL chip using PECVD. The tuning of the
QCL's single-mode lasing spectrum is shown as a function of the thickness of the deposited oxide. Spectra were measured
after each deposition step when the QCL was operated in a liquid-Nitrogen cooled dewar at $78~\rmK$ in pulsed mode at an
operating current of $\sim 570~$mA (current-density $\sim 405~\Appcm$). A net tuning of $\sim57~$GHz tuning is
demonstrated for an overall deposited thickness of $\sim3700~\rmnm$.
}
\label{Fig2}
\end{figure*}

A $10~\um$ thick terahertz QCL structure was grown by molecular beam epitaxy with a three-well resonant-phonon
GaAs/\AGAt\ design scheme~\cite{khanal:rpreview}. The metallic QCL cavities with antenna-feedback gratings were
fabricated using standard wafer-wafer thermocompression bonding and contact lithography technique.
Fig.~\ref{Fig1}(b) shows a 2D
finite-element (FEM) simulation of the dominant TM-polarized ($E_y$) electric-field for a typical cavity (of
infinite-width, for 2D simulation) with antenna-feedback gratings in the top-metal cladding. As can be seen, the the
resonant DFB mode excites a single-sided SPP wave in the surrounding medium, which is established as a standing-wave on
the top-metal cladding with a large spatial extent in the vertical dimension ($y$ direction).  Hence, any changes in the
refractive index of the surrounding medium serve to change the effective propagation constant $n_\rms$ of the
single-sided SPP mode, which tunes the excitation frequency of the resonant-DFB mode according to equation~\eqref{eq1}.


\section*{Results}

The active-medium of the QCLs is based on a three-well resonant-phonon design with GaAs/\AGAt\ superlattice (design
RTRP3W197, wafer number VB0464), which is described in Ref.~\cite{khanal:rpreview}, and was grown by molecular-beam
epitaxy.  The QCL superlattice is $10~\um$ thick with an average $n$-doping of $5.5\times 10^{15}~\iiicm$, and
surrounded by $0.05~\um$ and $0.1~\um$ thick highly-doped GaAs contact layers at $5\times 10^{18}~\iiicm$ on either side
of the superlattice. Fabrication of QCLs with parallel-plate metallic cavities followed a Cu-Cu thermocompression wafer
bonding technique as in Ref.~\cite{kumar:surfemit} with standard optical contact lithography.  Lateral and longitudinal
absorbing boundaries were implemented by exposing the highly doped GaAs layer in the finally fabricated cavities, and
the fabrication procedure is the same as in Ref.~\onlinecite{wu:feedback}. Ti/Au metal layers of thickness
$25/200~$nm were used as the top metal cladding, using an image-reversal lithography mask for implementing gratings in
the metal layers. Another positive-resist lithography step was used to cover the grating-metal with photoresist to be
used as a mask for wet-etching of ridges in a H$_2$SO$_4$:H$_2$O$_2$:H$_2$O $1$:$8$:$80$ solution. A Ti/Au contact was
used as the backside-metal contact for the finally fabricated QCL chips to assist in soldering.  Before deposition of
backside-metal of the wafer, the substrate was mechanically polished down to a thickness of $\sim 170~\um$ to improve
heat-sinking.

By deposition of Silicon-dioxide post-fabrication (after the QCL chip is already mounted and wire-bonded on a copper
mount), the lasing frequency of the resonant antenna-feedback mode is tuned sensitively. Fig.~\ref{Fig2} shows the
variation of a specific QCL's wavelength of emission when increasing the thickness of blanket deposited Silicon-dioxide
on the entire QCL chip. As expected from equation~\eqref{eq1}, the lasing wavelength undergoes a red-shift with
increasing thickness of the oxide, since then, the effective propagation index $n_\rms$ for the SPP wave in the
surrounding medium of the QCL will increase. Additional details about changes to the hybrid SPP mode, and 
correspondingly $n_\rms$, with oxide deposition are provided in supplementary material (section S1).
The measured QCL has a grating period $\Lambda = 25~\um$ and the cavity's
dimensions are $100~\um \times 1.4~\rmmm\times 10~\um$. A cleaved chip consisting of QCLs with antenna-feedback gratings
was In-soldered on a Cu block, the QCL to be tested was wire-bonded for electrical biasing, and the Cu block was mounted
on the cold-stage of a liquid Nitrogen vacuum cryostat for measurements. In this experiment, the frequency of the QCL
was first measured without any dielectric deposition, and it radiated in a single-mode at $\sim2.85~\thz$ ($\lambda\sim
105~\um$). Afterwards, the resonant-frequency is changed by depositing Silicon-dioxide on top of the mounted QCL chip
using a plasma-enhanced chemical vapor deposition (PECVD) system. The deposited thickness is carefully calibrated by
simultaneously including a bare GaAs wafer in the PECVD chamber for each deposition step. QCL's spectra were measured at
$78~\rmK$ in linear-scan mode with a resolution of $0.2~\icm$ using a FTIR with room-temperature pyroelectric detector.
Multiple cycles of spectral measurement and oxide-deposition were implemented and the tuning results are collected and
shown in Fig.~\ref{Fig2}, which illustrates a large red-shift tuning with this mechanism. Finally, a mode-hop-free and
continuous tuning of $\sim 57$ GHz was achieved after an overall deposited thickness of $\sim3700~\rmnm$. Further tuning
could not be realized since QCL stopped lasing at $78~\rmK$ beyond the stated thickness. The spectra shown in
Fig.~\ref{Fig2} were all measured at similar bias currents. When etching away the deposited SiO${_2}$ on top of the
laser device by buffered-HF (BOE $7:1$) etchant, the QCL was characterized again and its emission properties including the
lasing wavelength returned to its original value, which illustrates that tuning by post-process deposition of SiO${_2}$
on terahertz QCLs here is a reversible method.  

\begin{figure*}[htbp]
\centering
\includegraphics[width=5.0in]{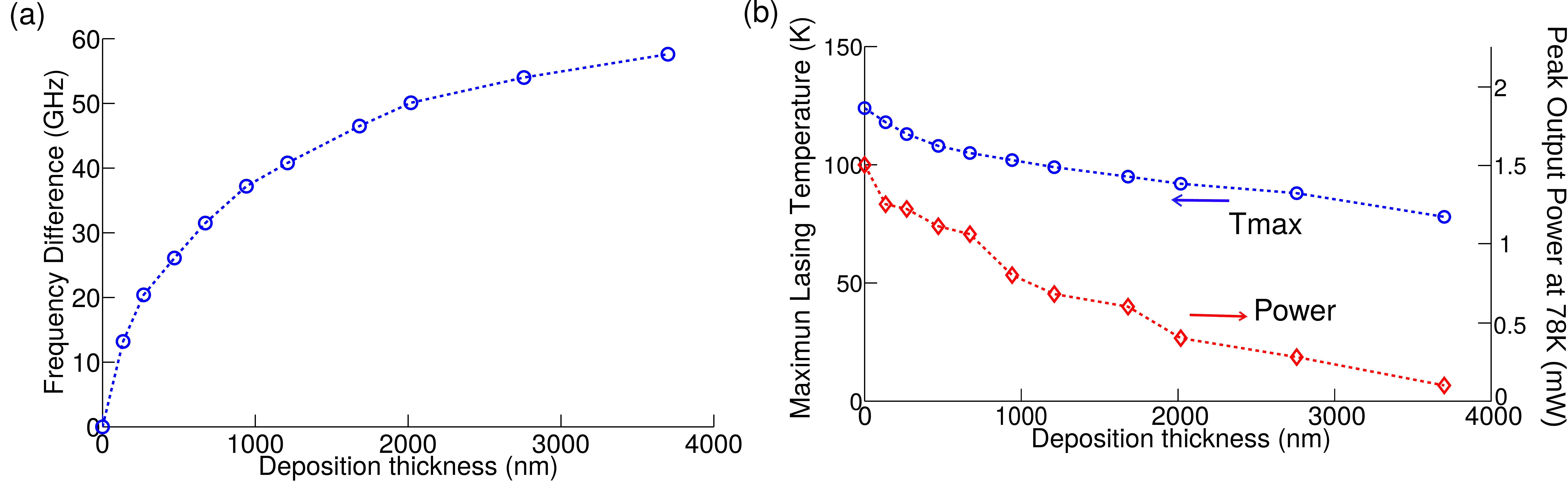}
\caption{
(a)~Emission frequency of the terahertz QCL with antenna-feedback as a function of the deposited thickness of
Silicon-dioxide, expressed as a difference from its original emission frequency of $\sim 2.85~\thz$ without any oxide.
The QCL is biased slightly below the peak-power bias region where it radiates in a single-mode (detailed spectra with
bias are shown in Fig.~\ref{Fig4}).  (b)~Maximum operating temperature ($\Tmax$) and the detected peak optical power at
$78~\rmK$ for the QCL in pulsed operation, as a function of the thickness of the Silicon-dioxide.
}
\label{Fig3}
\end{figure*}

\begin{figure*}[htbp]
\centering
\includegraphics[width=5.0in]{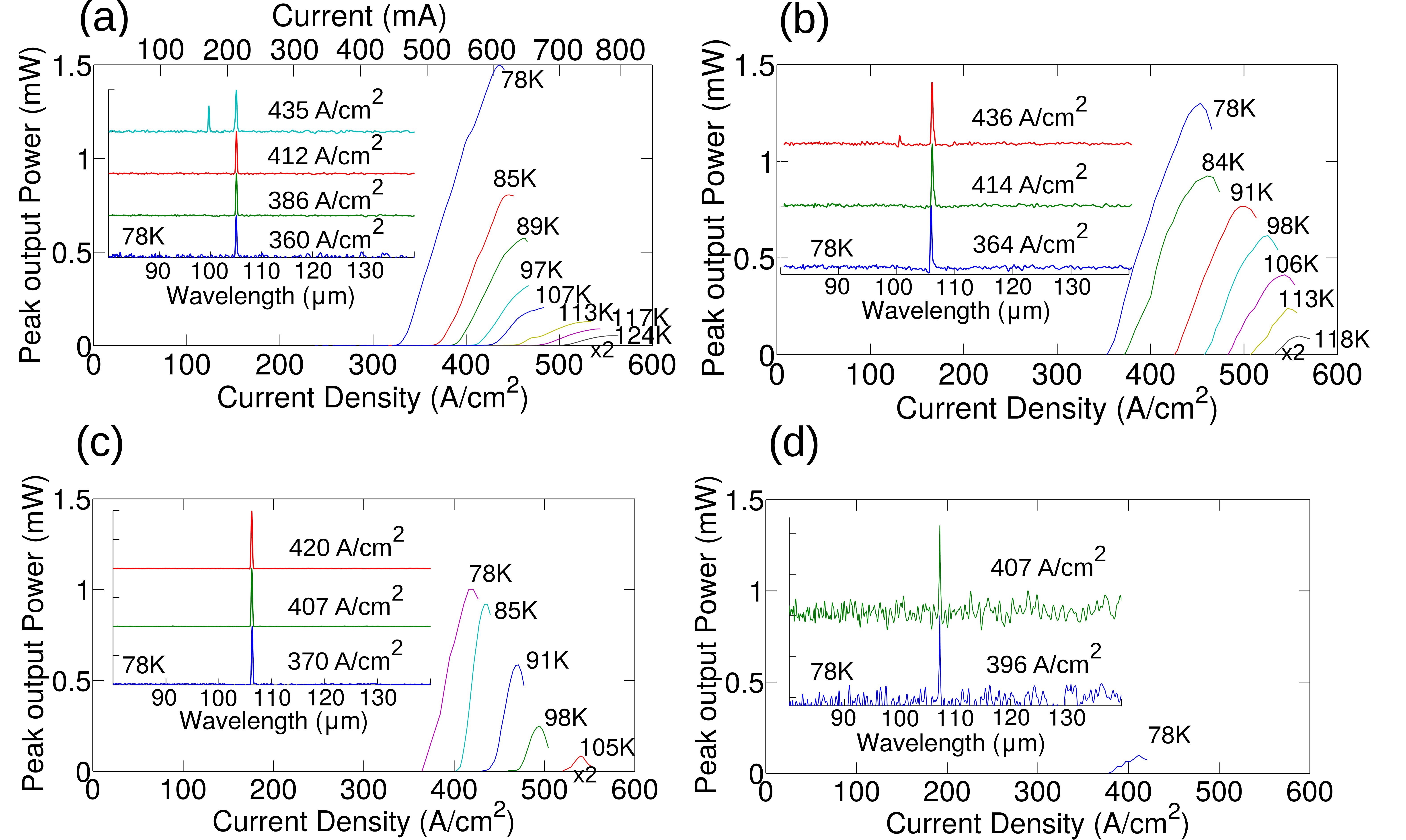}
\caption{
Light-current characteristics of the terahertz QCL with antenna-feedback scheme at different heat-sink
temperatures.  The dimensions of the QCL's cavity are $100~\um\times 1.4~\rmmm\times 10~\um$. The QCL is biased in
pulsed mode with $400~$ns wide pulses repeated at $100~$kHz. 
The plot-insets show lasing spectra at $78~\rmK$ as a function of increasing bias. (a)~Original QCL without any oxide
deposition, (b)~QCL with $140~$nm thick oxide leading to $\sim13~\ghz$ frequency tuning, (c)~QCL with $700~$nm thick
oxide leading to $\sim31~$GHz frequency tuning, and (d)~QCL with $3700~$nm thick oxide leading to the maximum tuning of
$\sim57~$GHz for the QCL still lasing at $78~\rmK$.
}
\label{Fig4}
\end{figure*}

Figure~\ref{Fig3}(a) characterizes the variation of the QCL's emission frequency as a function of thickness of the
deposited Silicon-dioxide. The variation is non-linear, and the rate of tuning as a function of oxide-thickness
decreases as more oxide is deposited. A 2D finite-element simulation was not able to reproduce this non-linear tuning
since the hybrid SPP mode in the surrounding medium of the cavity in the vertical direction (above the cavity) has a
large spatial extent. Correspondingly, the tuning computed with 2D simulations is predominantly linear for an overall
oxide thickness of few microns. However, it is argued that the non-linear tuning behavior as observed here is due to the
complex spatial nature of the SPP mode in the surrounding medium in the lateral dimensions~\cite{wu:feedback}, which
is difficult to capture even with 3D simulation due to hardware limitations in
performing such simulations. The experimental tuning result is compared to that predicted from both 2D and 3D simulations
in the supplementary material (section S2). The dependency of measured maximum lasing temperature and peak output power at
$78~\rmK$ with oxide thickness is shown in Figure~\ref{Fig3}(b). Without any post-process deposition of the oxide, this
terahertz QCL device lased up to $124~\rmK$ with peak output-power of $\sim 1.5~\rmmW$ at $78~\rmK$. When deposition
thickness reaches $\sim3700~\rmnm$ and $\sim 57$ GHz tuning is obtained, the maximum lasing temperature of this device
is slightly above $78~\rmK$ and the peak output-power reduces to $\sim 100~\rmu$W at $78~\rmK$. The emitted optical power
was measured with a deuterated triglycine sulfate pyroelectric detector (DTGS) pyroelectric detector (Gentec THz
2I-BL-BNC) and calibrated with a terahertz thermopile power-meter (ScienTech model AC2500H) without any optical component
or cone collecting optic inside the cryostat. The power values are reported without any corrections to the detected
signal.

Two reasons contribute to the decrease in maximum lasing temperature and output-power of the QCL with increasing oxide
thickness. First, thin-film oxide deposited by PECVD is a lossy material at terahertz frequencies for which the
absorption coefficient could be as high as $20~\icm$~\cite{lee:tera}. Secondly, the confinement factor $\Gamma$ of the
resonant-cavity mode, which is the fraction of the mode that resides in the active-medium, decreases as thicker oxide is
deposited. This is primarily because the higher electric-permittivity of the oxide $\epsilon_\mathrm{ox}\sim
4.4\epsilon_0~$ (at terahertz frequencies). As a consequence the effective threshold gain is increased for exciting the
resonant-cavity mode, which also reduces optical power due to an increase in the effective modal propagation loss in the
cavity as well as reduction in the net dynamic range for lasing current. A precise analytical or numerical estimation
of the net effect of oxide thickness on the threshold gain and QCL's temperature and power performance is beyond the
scope of this work because, first, in case of terahertz QCLs various transport parameters as well as waveguide loss parameters 
cannot be predicted accurately, and second, the estimation of $\Gamma$ requires 3D finite-element simulation of the 
QCL's cavity with deposited oxide that cannot be done due to aforementioned reasons.

\begin{figure*}[htbp]
\centering
\includegraphics[width=5.0in]{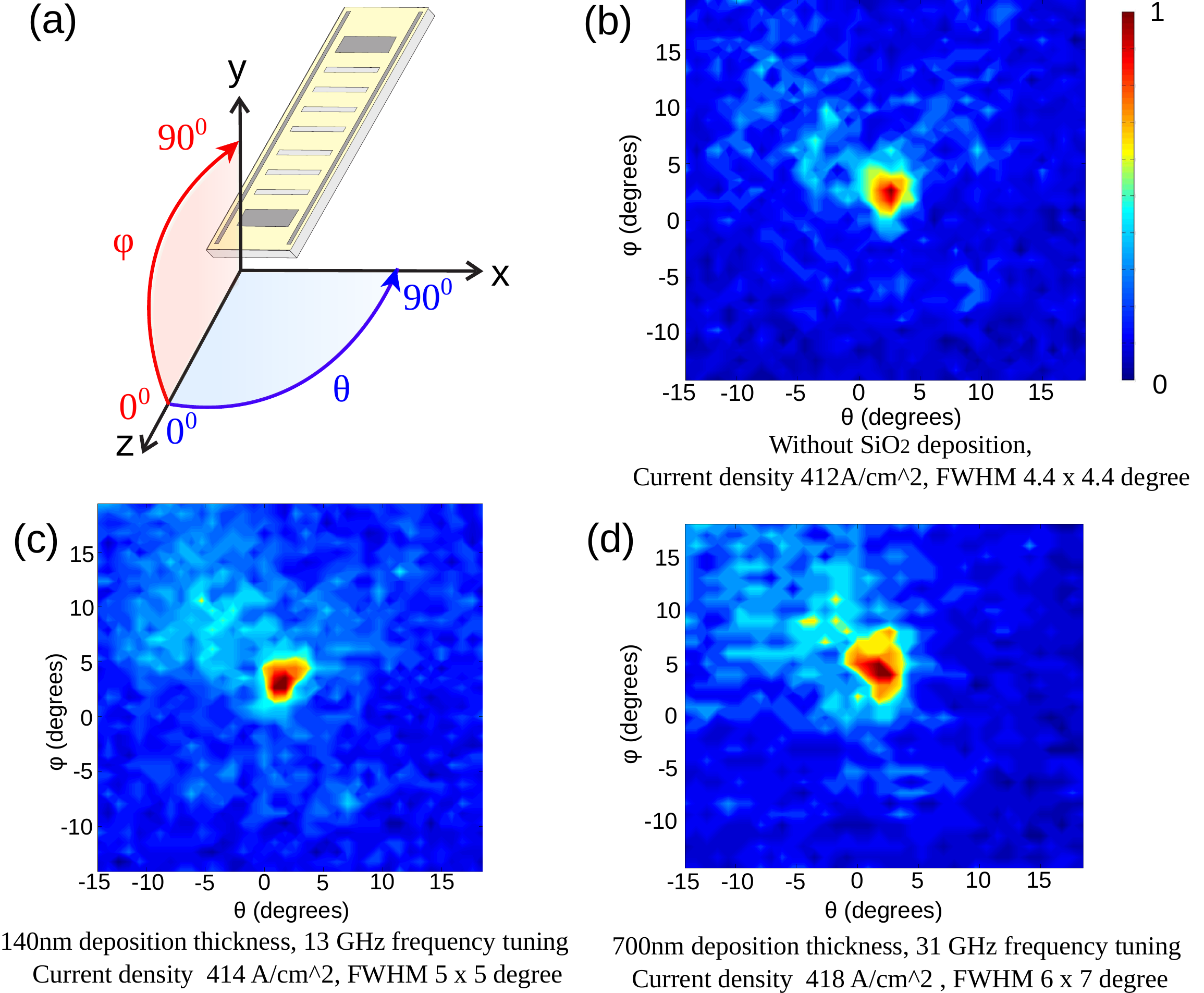}
\caption{
(a)~Schematic showing orientation of the QCL and definition of angles for the shown far-field radiation patterns.
The QCL was operated at $78~\rmK$ in pulsed mode, and the radiation pattern was measured using a room-temperature
pyroelectric detector of $2~$mm diameter mounted on a $x-y$ movement stage. The radiation-pattern in
(b) is for QCL without Silicon-dioxide deposition, in (c) is for QCL deposited with an oxide thickness of $140~$nm, and in (d)
is for an oxide thickness of $700~$nm.
}
\label{Fig5}
\end{figure*}

Figure~\ref{Fig4} shows light-current (\LI) characteristics in pulsed operation for the QCL operated at different
heat-sink temperatures along with its lasing spectra at different bias when varying deposition thickness of the
Silicon-dioxide. 
The QCL predominantly radiates at a single frequency around $2.84~\thz$ that corresponds to the desired DFB mode of the
cavity. However, at higher current-densities closer to peak operating bias ($>420~\Appcm$), a second mode at
shorter-wavelength is excited, which suggests that a higher-order lateral mode was likely excited due to spatial-hole
burning in the cavity (since such a mode has a smaller effective propagation index $n_\rmc$ in the cavity).  The
excitation of the higher-order lateral mode ceases in the entire operating range of the QCL as thicker oxide is
deposited.  When the oxide thickness is increased to $\sim700~\rmnm$ corresponding to $\sim31~$GHz frequency tuning, the
QCL lased up to $105~\rmK$ with peak output-power of $\sim 1~\rmmW$ at $78~\rmK$. Single mode spectra with lasing
frequency $\sim2.82~\thz$ were measured under all bias conditions over the whole dynamic range, namely from threshold
until negative differential-resistance (NDR) as shown in Figure~\ref{Fig4}(c). When total thickness of the deposited
oxide is $\sim 3700~\rmnm$, only limited spectra were recorded at current-densities of $\sim 396~\Appcm$ and $\sim
407~\Appcm$ respectively while the QCL is still operating at $78~\rmK$ in a liquid-Nitrogen cooled dewar. The QCL
stopped lasing at $78~\rmK$ when additional oxide was deposited. The threshold current-densities at $78~\rmK$ for the
aforementioned plots as shown in Fig.~\ref{Fig4} are $\sim 328~\Appcm$, $\sim 350~\Appcm$, $\sim 364~\Appcm$, $\sim
390~\Appcm$ respectively, which increase with the thickness of the deposited oxide. From the data presented in
Fig.~\ref{Fig3} and Fig.~\ref{Fig4} it can be noted that more than $60~\%$ of the available tuning (\ie\ $\sim 35~\ghz$
tuning) could be realized by depositing less than $1~\um$ thick Silicon-dioxide, in which case there is no significant
degradation in maximum operating temperature or power output from the QCL. 

Radiation in a directional (narrow) beam is one of the most important and attractive characteristics of terahertz
plasmonic QCLs with the antenna-feedback scheme~\cite{wu:feedback}.  Figure~\ref{Fig5} shows measured beam patterns of
the QCL at $78~\rmK$ for varying thicknesses of the deposited oxide. Far-field radiation patterns were measured with a
DTGS pyroelectric cell detector mounted on a computer-controlled two-axis moving stage in the end-fire ($z$) direction
at a distance of $65~$mm from the QCL's end-facets. Without any oxide, the QCL emits in a single-lobed beam with narrow
divergence with full-width half-maximum (FWHM) of $\sim 4.4^{\circ} \times 4.4^{\circ}$ as shown in Fig.~\ref{Fig5}(b).
After $140~$nm thick oxide was deposited, the radiation pattern remained single-lobed but with a slightly increased FWHM
of $\sim 5^{\circ} \times 5^{\circ}$. A further deposition of oxide to a thickness of $700~$nm increased the FWHM of the
main lobe to $\sim 6^{\circ} \times 7^{\circ}$. The small increase in beam divergence for thicker oxide deposition is
likely due to reduction in spatial extent of the SPP mode on top of the cavity with oxide thickness.  Due to the limits
in the sensitivity of the of the room-temperature DTGS detector, the radiation patterns could not be measured with a
reasonable signal-to-noise ratio after deposition of even thicker oxide. However, since the QCL continues to radiate in
the same cavity mode, it should retain its characteristic radiation into single-lobed narrow beams.

\section*{Discussion}

Plasmonic lasers utilize metallic cavities to confine the electromagnetic mode at subwavelength dimensions. However,
poor radiative coupling to free-space makes it difficult to statically or dynamically tune their emission frequency,
which is required for practical applications. The tuning mechanisms demonstrated so-far involve alteration of
gain-medium itself. Terahertz QCLs with metallic cavities suffer from similar challenges that has prevented development
of tunable QCLs that could work at high-temperatures. In this work, we show that a recently developed antenna-feedback
scheme for plasmonic lasers could offer an ideal solution for both static tuning (as demonstrated here) or dynamic
tuning and frequency modulation (in principle, which is discussed further in section S3 of the supplementary material),
while simultaneously achieving single-mode operation and narrow-beam
emission from the sub-wavelength plasmonic cavities.  The antenna-feedback scheme leads to the establishment of a hybrid
SPP standing-wave in the surrounding medium of the laser's cavity with a large spatial extent. The emission frequency of
the plasmonic laser depends sensitively on the effective propagation index of the SPP mode in the surrounding medium,
which could be altered independently of the gain medium used inside the laser's cavity. Consequently small
perturbations in the refractive-index of the surrounding medium could lead to large modulation in the laser's emission
frequency.  To the best of our knowledge, this is a unique mechanism to tune the frequency of a solid-state laser for
which there is no analogous precedent in prior literature.

For terahertz QCLs, the antenna-feedback scheme was implemented for wide cavities that could operate well above the
temperature of liquid-Nitrogen, in contrast with previous tuning techniques that have mostly utilized ultra-narrow
cavities for access to the evanescent cavity modes. $57~\ghz$ static tuning is demonstrated for single-mode narrow-beam
terahertz QCLs emitting at $2.8~\thz$, based on post-process deposition of Silicon-dioxide on an already soldered and
mounted QCL chip. A beam divergence of $<7^\circ$ was maintained through the tuning range. The degradation of maximum
operating-temperature and power output is small for a large fraction of the tuning range.  More than $50~\%$ of the
available tuning ($\sim 30~\ghz$ tuning) could be realized by depositing $\sim 0.7~\um$ thick Silicon-dioxide, in which
case $\Tmax$ reduces from $124~\rmK$ to $105~\rmK$ and the peak output-power reduces from $1.5~\rmmW$ to $1~\rmmW$.  The
tuning is reversible and continuous, and could pave the way for future commercialization of DFB terahertz QCLs for
targeted applications in sensing and high-resolution spectroscopy. Whereas all previous tuning results for terahertz
QCLs were demonstrated at operating temperatures close to that of liquid-Helium, the results presented here are for QCLs
operating at the much more practical temperature of $78~\rmK$, while simultaneously achieving significantly more directional
beams compared to previous reports. 

\section*{Acknowledgments}
This work is supported by the United States National Science Foundation through grants ECCS 1351142 and CMMI 1437168.
The work is performed, in part, at the Center for Integrated Nanotechnologies, a U.S. Department of Energy (DOE), Office
of Basic Energy Sciences user facility. Sandia National Laboratories is a multiprogram laboratory managed and operated
by Sandia Corporation, a wholly owned subsidiary of Lockheed Martin Corporation, for the U.S. DOE’s National Nuclear
Security Administration under contract DE-AC04-94AL85000.


\section*{Supplemental Documents}
See supplementary material for discussions about the effective propagation index of the hybrid SPP wave in the
surrounding medium, finite-element simulations of the tuning behavior, and a proposed dynamic tuning scheme.


%

\newpage
\title{Large static tuning of narrow-beam terahertz plasmonic lasers operating at $78~\rmK$}

\author{Chongzhao Wu}
\email{chw310@lehigh.edu}
\author{Yuan Jin}%
\author{John L. Reno}
\affiliation{
Sandia National Laboratories, Center of Integrated Nanotechnologies,
MS 1303, Albuquerque, NM 87185-1303\\
}
\author{Sushil Kumar}%
\affiliation{%
Department of Electrical and Computer Engineering, Lehigh University, Bethlehem, PA 18015, USA\\
}%

\date{\today}


\begin{center}
{\Large Large static tuning of narrow-beam terahertz plasmonic lasers operating at $78~\rmK$}
\\
\vspace{0.2in}
Chongzhao Wu, Yuan Jin, John L. Reno and Sushil Kumar
\\
\vspace{0.4in}
{\Large Supplementary Information}
\\
\vspace{0.1in}
\end{center}


\section{Effective propagation index of the hybrid SPP mode}

The wavelength selected by antenna-feedback scheme is $\lambda = \Lambda
(n_\rmc + n_\rms)$ as shown in Ref.~\cite{wu:feedback}, which is different from
all previously reported $p-th$ order distributed-feedback (DFB) schemes in
which the DFB mode occurs at  $\lambda = 2 n_\rmc \Lambda/p$.
Here, $n_\rms$ is an ``effective'' propagation index for the hybrid SPP wave
propagating in the surrounding medium of the cavity that is bound to the top
metallic cladding. It plays characteristic role in determining the resonant
frequency of DFB mode and its tunability, since the wavelength of the resonant-cavity
mode depends sensitively on $n_\rms$. It is beyond the scope of this work
to describe the spatial profile of the electric-field corresponding to the
surrounding SPP wave analytically, even for the relatively simple case when no
SiO$_2$ is present.  In the following, the profile of the SPP wave is computed
numerically via finite-element simulations.  As expected, $n_\rms$ is closer to
that of SiO$_2$ as the thickness of SiO$_2$ is increased, since a larger
fraction of the SPP wave will then exist in SiO$_2$ instead of air (vacuum).

\begin{figure*}[htbp]
\centering
\includegraphics[width=6in]{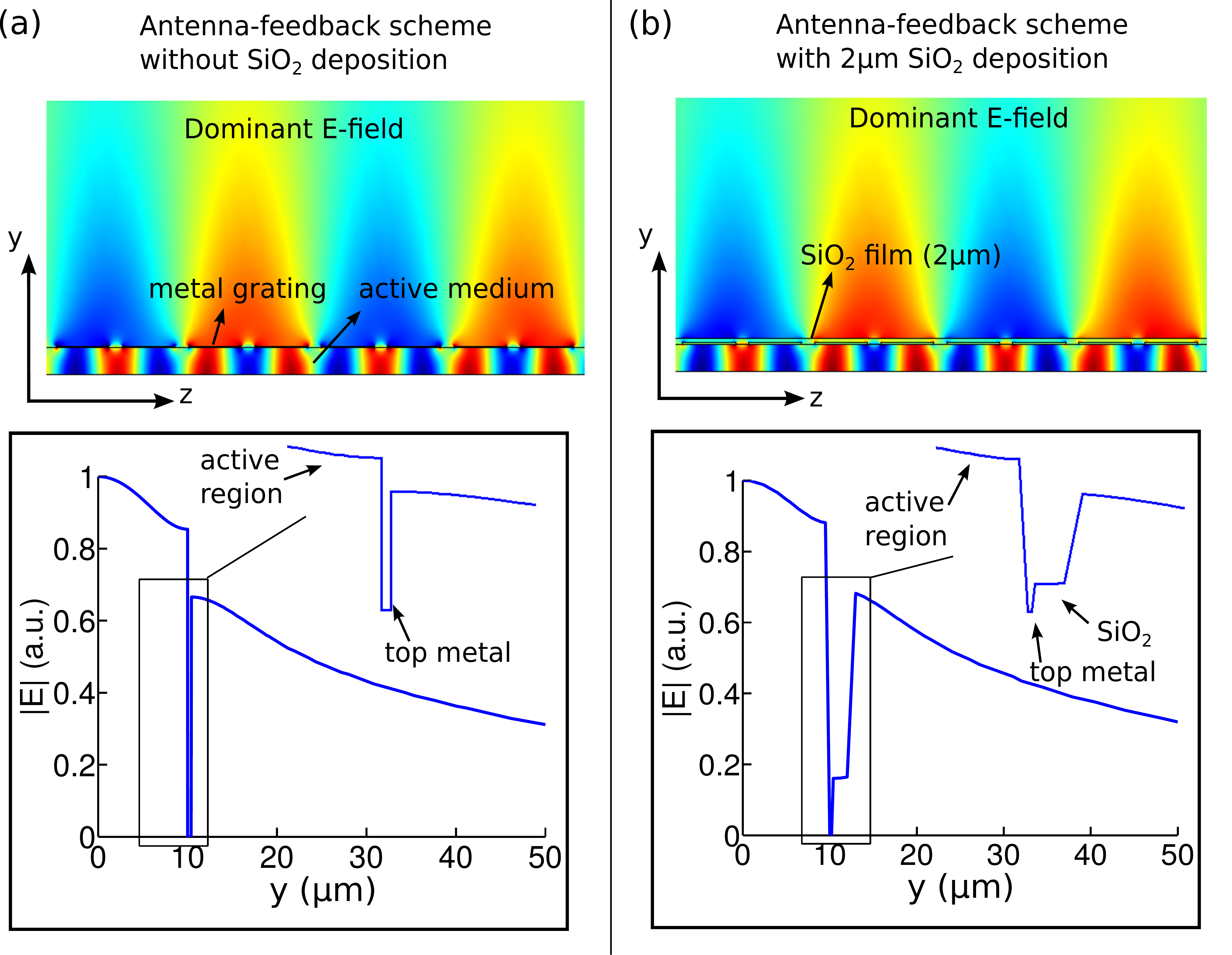}
\caption{
{\bf Comparison of the electric-field profile of the resonant-cavity mode with and without
SiO$_2$ deposition}
(a)~Magnitude of the electric-field $|E|$ for the lowest-loss resonant-cavity mode
computed by 2D finite-element simulations. The $10~\um$ thick GaAs cavity has antenna-feedback
gratings in its top metal cladding. The mode-confinement factor for the active-region is $\sim 70~\%$.
(b)~Electric-field profile for the same cavity but now with a $2~\um$ thick SiO$_2$ layer added on top of the
cavity. In this case, $\sim 1~\%$ of the mode's energy is confined in the SiO$_2$ layer, which is approximately
$4~\%$ of the energy in the hybrid SPP wave outside the cavity/active-region.
}
\label{figS1}
\end{figure*}

Fig.~\ref{figS1}(a) shows the absolute value of electric-field, $|E|$, both
inside the cavity and above the metal cladding for the antenna-feedback scheme
from a 2D simulation (i.e. for a cavity with 'infinite' lateral width). The
layer sequence includes an active region of $10~\um$ thickness and top metal
cladding of $400~$nm thickness that is modeled as lossless metal. The
sub-wavelength plasmonic resonant optical mode inside the cavity is
phased-locked with a hybrid SPP mode that travels on top of the metal cladding
outside the cavity as mentioned in the main text. The SPP wave is relative
tightly bound to the top metal cladding and decays (somewhat) exponentially in
the length of the order of the wavelength in the surrounding material/medium.
Attributed to the fact that the decay length in metal is negligible and
independent of the loss in metal or thickness in metal (which was verified
independently via finite element simulations with lossy metal), it is possible
to qualitatively estimate the effect of SiO$_2$ film on the effective refractive
index $n_\rms$ of single-side SPP mode that propagates on the top of the cavity.

Fig.~\ref{figS1}(b) shows the simulation result for a specific example in which
the deposited SiO$_2$ film is $2~\um$ thick on the top of the laser cavity. The
electric-energy stored in SiO$_2$ layer is computed to be $\sim 4~\%$ of the
total energy in the SPP wave in the surrounding medium of the cavity, which
leads to the effective $n_\rms$ becoming $\sim 1.045$
($=0.04n_\mathrm{oxide}+0.96n_\mathrm{air}$) calculated by the weighted
contributions from SiO$_2$ and vacuum according to the fraction of confined
energy in the respective medium (where the refractive index of SiO$_2$ is taken as
$n_\mathrm{oxide}\sim 2.1$ at terahertz frequencies). This provides a general
idea about the change in resonant-frequency of the DFB mode, since from eq.~(1)
in the main text,
\begin{equation}
-\frac{d \nu}{\nu} = \frac{d \lambda}{\lambda} = \frac{d n_\rms}{n_\rms} \left(\frac{n_\rms}{n_\rmc+n_\rms}\right)
\nonumber
\end{equation}
An increase in $n_\rms$ by $\sim 4.5~\%$ should lead to a decrease of resonant
frequency by $\sim 1~\%$ according to the expression above, which is indeed
observed from the simulated results for the resonant-frequency of the DFB mode.
This verifies the validity of eq. (1) from the main text and the fact that $n_\rms$ is an effective
propagation index of the hybrid SPP wave. However, note that the analysis based
on 2D simulations will not provide an accurate estimate of the effective index
$n_\rms$ for real experiments. The SPP wave in the surrounding medium extends
significantly in the lateral directions beyond the width of the cavity as
described in Ref.~\cite{wu:feedback}, and hence attains a complex
3D profile. Even small variations in the geometry of the cavity (including the
angle of the sidewalls and the distance to the neighboring cavities) affects
the shape of the hybrid SPP wave, and correspondingly the resonant-frequency of
the DFB mode. Hence, the simplistic 2D finite-element simulations could be
utilized to understand the modal behavior only qualitatively.

\begin{figure*}[htbp]
\centering
\includegraphics[width=6in]{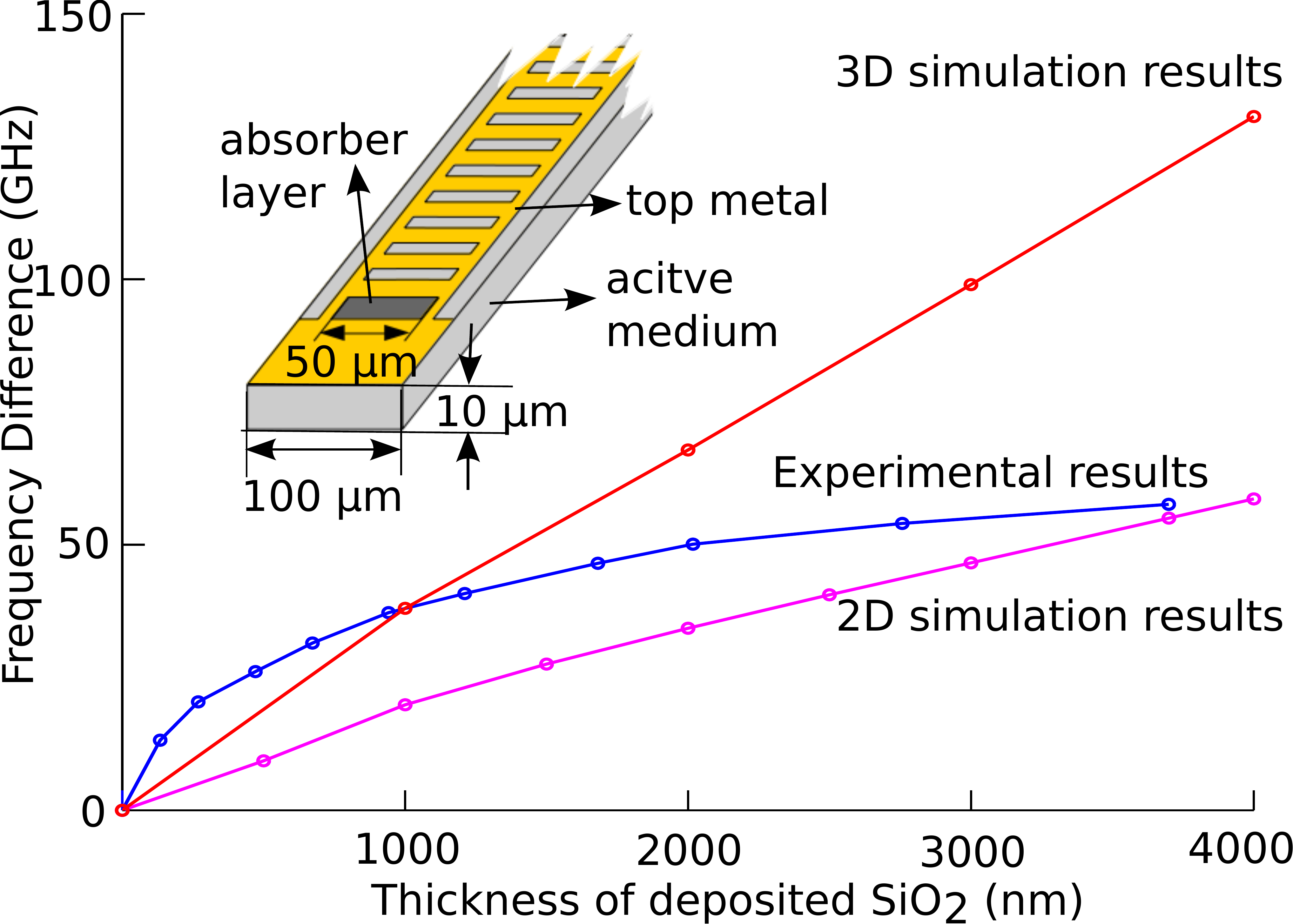}
\caption{
{\bf Frequency tuning versus deposited oxide thickness for the lowest-loss resonant-cavity
mode for a QCL cavity with antenna-feedback scheme}
Experimental tuning result for a $10~\um$ thick
cavity is plotted alongside that obtained through 2D simulations (that models a cavity with infinite
width), as well as results from full-wave 3D simulation in which the shown geometry is modeled for a
cavity of $1.4~\rmmm$ length.
}
\label{figS2}
\end{figure*}

\section{Finite-element simulation of tuning characteristics}

A comparison between simulated and experimental tuning curves as a function
of the thickness of deposited oxide is shown in Fig.~\ref{figS2}.
The tuning computed with 2D simulations is predominantly linear for an overall
oxide thickness of few microns. However, it is argued that
the non-linear tuning behavior as observed in experiments is due to the complex
spatial nature of the hybrid SPP mode in the surrounding medium for cavities
with finite width, which was also discussed in Ref.~\cite{wu:feedback}. The
modal behavior can only be captured partially by 3D simulations due to
limitations in accuracy of the modeled geometry as well as memory limitations
for finite-element simulations that require the absorbing boundaries to be
placed very close to the cavity for simulation. The thin-film oxide deposited
by PECVD is a lossy material at terahertz frequencies. However, the accurate
modeling of loss due to oxide is beyond the scope of this work, especially since
the loss-parameters of the low-temperature ($\sim \degC{100}$) deposited oxide are
not known.

To further explain the non-linear tuning performance, some 3D
simulation results are also included in the Fig.~\ref{figS2}. Considering the
limitations in computer memory, the resolution of photolithography, and other
uncertainties in fabrication (that include the curved sidewall profiles of the
wet-etched cavities that is difficult to model), some simplifications are made
in the creation of the geometry of 3D module. From the simulation results shown
in Fig. S2, the tuning performance can be seen to vary significantly for 2D versus
3D simulation. It was also noticed that the simulated results were different if the
3D geometry was altered. Also, the effect of neighboring cavities cannot be
captured in such a simulation due to lack of computer memory to simulate
multiple cavities. Due to these reasons, any other simulations to estimate the
non-linear tuning effect have proven to be futile.

It should also be noted that the large non-linear effect of tuning might also
be attributed to some other reasons that may be difficult to model. One factor
that may affect tuning performance is the quality of the deposited oxide.
Considering the Indium-solder mounted QCL chip in our experiments, a low
deposition temperature ($\sim \degC{100}$) was chosen for deposition, which
resulted in poor quality of the deposited oxide. SEM images of the deposited
SiO$_2$ show cracks on the surface of the SiO$_2$ (not shown here). The effect
of oxide quality on tuning behavior is difficult to envision. However, we could
safely conclude that the temperature performance of QCLs with deposited oxide
could be improved if a better quality oxide can be deposited.

\begin{figure*}[htbp]
\centering
\includegraphics[width=6in]{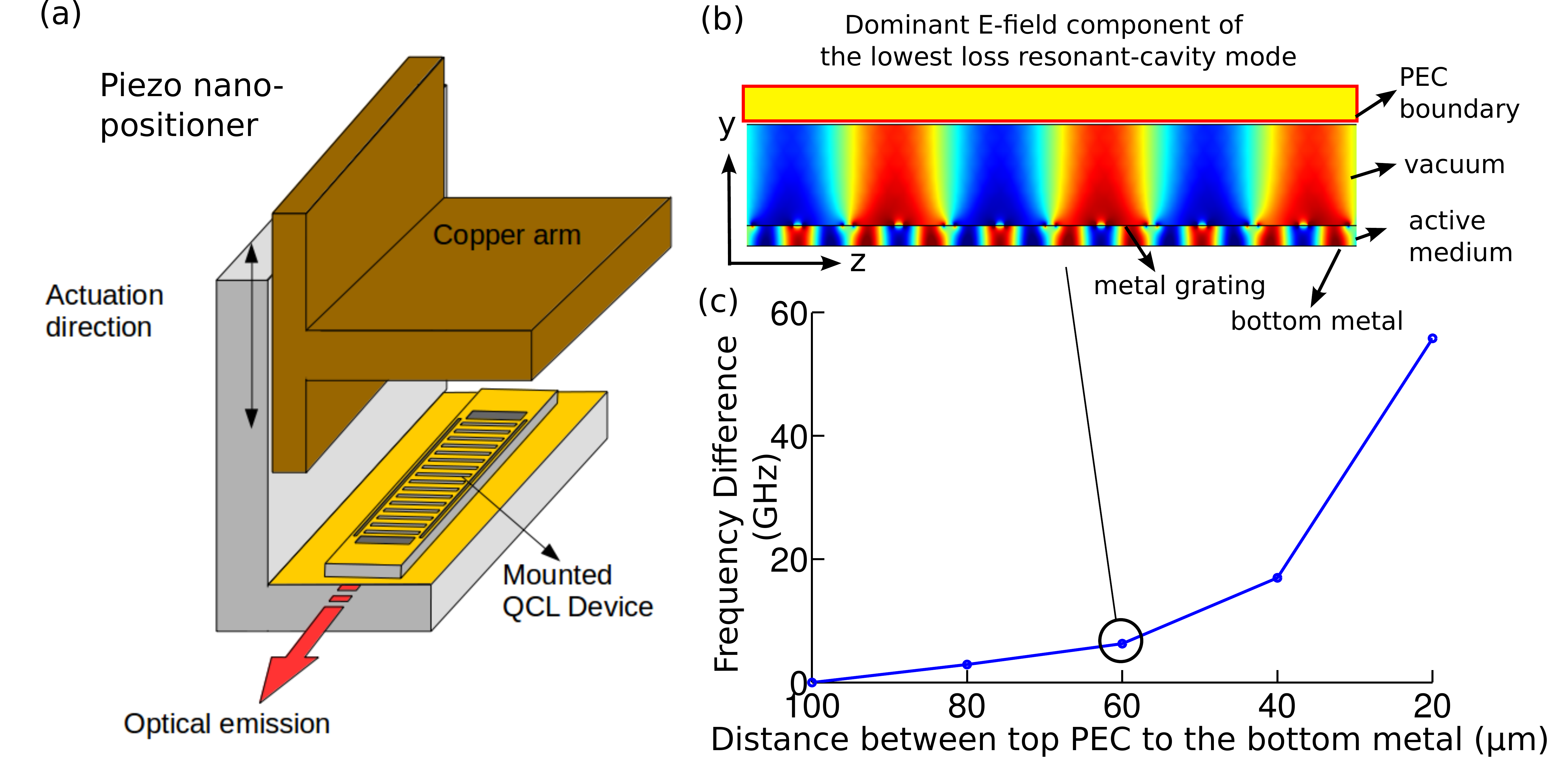}
\caption{
{\bf Illustration of a scheme to dynamically tune the resonant-cavity mode of a cavity with
antenna-feedback scheme based on an electrically movable nano-positioner}
(a)~A possible
dynamic tuning setup by using a metallic arm controlled by an piezo nano-positioner, located over the
QCL cavity with antenna-feedback. (b)~Mode-shape of the resonant-cavity mode with the metallic arm
in vicinity of the cavity in the surrounding medium, modeled as perfect-electric-conductor in 2D finite-
element simulations. (c)~Frequency tuning results from the 2D simulation for the lowest-loss resonant-cavity
mode, computed as a function of distance of the metallic arm from the bottom metallic plane of
the cavity.
}
\label{figS3}
\end{figure*}

\section{Dynamic tuning based on an electromechanical device}

A relatively straightforward method to achieve dynamic tuning of terahertz QCLs
with antenna-feedback scheme would be to utilize electromechanical
nano-positioners, similar to those utilized for some terahertz QCLs in prior
work as described in Refs.~\cite{qin:tunableMEMS,han:broad,castellano:tuning}.
Thanks to the establishment of the strong hybrid SPP mode on the top of the
cavity in vacuum, large dynamic tuning is predicted based on simulations for
such cavities with antenna-feedback scheme. The primary advantage of such a
tuning scheme would be the capability to use wider cavities that will have a
much better temperature performance, while still retaining narrow far-field
radiation patterns are are a characteristic of the antenna-feedback scheme.

Fig.~\ref{figS3} illustrates a piezoelectric nano-positioner that could be
utilized to control the distance between the laser cavity and an overhanging
metallic arm attached to the nano-positioner. A 2D simulation was carried out
to estimate the tuning performance for such an implementation. In this
simulation, the metal arm was modeled as a perfect-electric-conductor (PEC)
boundary. When the distance between the bottom metal of the laser cavity to the
PEC boundary is changed from $100~\um$ to $40~\um$, $\sim 17~\ghz$ tuning is estimated for
the resonant-cavity DFB mode. The SPP wave in the surrounding medium is
considerably altered when the PEC boundary gets even closer and a large tuning
of $\sim 60~\ghz$ is predicted if the metallic arm could be brought to a distance of
$\sim 20~\um$ from the ground plane, which should be realizable for piezo-based
positioners with positioning accuracy of few microns.

\end{document}